\documentclass[a4paper,11pt]{article}
\pdfoutput=1 

\usepackage{jinstpub} 
\usepackage{lineno,hyperref}
\usepackage{textcomp}
\usepackage[usenames,dvipsnames]{color}

\title{Gas Distribution and Monitoring for the Drift Chamber of the MEG-II Experiment}

\author[a]{A.~M.~Baldini,}
\author[b,1]{E.~Baracchini,\note{Now at Gran Sasso Science Institute, Via Francesco Crispi 7, 67100 L'Aquila, Italy}}
\author[c,d]{G.~Cavoto,}
\author[a,e]{F.~Cei,}
\author[a,f]{M.~Chiappini,}
\author[c]{G.~Chiarello,}
\author[g,h]{C.~Chiri,}
\author[a,e]{M.~Francesconi,}
\author[a]{L.~Galli,}
\author[g]{F.~Grancagnolo,}
\author[a]{M.~Grassi,}
\author[i]{M.~Hildebrandt,}
\author[l]{V.~Martinelli,}
\author[c,d]{M.~Meucci,}
\author[a,e]{D.~Nicol\`o,}
\author[g,h]{M.~Panareo,}
\author[i]{A.~Papa,}
\author[g,h]{A.~Pepino,}
\author[a,e]{B.~Pruneti,}
\author[a]{F.~Raffaelli,}
\author[c,2]{F.~Renga,\note{Corresponding author}}
\author[c,3]{E.~Ripiccini,\note{Now at Universit\'e de Gen\'eve D\'epartement de physique nucl\'eaire et corpusculaire, Quai Ernest-Ansrmet 24, 1205 Gen\'eve, Switzerland}}
\author[a]{G.~Signorelli,}
\author[g,h]{G.~F.~Tassielli}
\author[c]{and C.~Voena}

\affiliation[a]{INFN Sezione di Pisa, Largo B.~Pontecorvo~3, 56127 Pisa, Italy }
\affiliation[b]{ICEPP, The University of Tokyo, 7-3-1 Hongo, Bunkyo-ku, Tokyo 113-0033, Japan}
\affiliation[c]{INFN Sezione di Roma, Piazzale A.~Moro 2, 00185 Roma, Italy} 
\affiliation[d]{Dipartimento di Fisica dell'Universit\`a ``Sapienza'' di Roma, Piazzale A.~Moro 2, 00185 Roma, Italy}
\affiliation[e]{Dipartimento di Fisica dell'Universit\`a di Pisa, Largo B.~Pontecorvo~3, 56127 Pisa, Italy}
\affiliation[f]{Dipartimento di Scienze Fisiche, della Terra e dell'Ambiente dell'Universit\`a di Siena, Via Roma 56, 53100, Siena, Italy}
\affiliation[g]{INFN Sezione di Lecce, via per Arnesano, 73100 Lecce, Italy} 
\affiliation[h]{Dipartimento di Matematica e Fisica dell'Universit\`a del Salento, Via per Arnesano, 73100 Lecce, Italy} 
\affiliation[i]{Paul Scherrer Institut PSI, 5232 Villigen, Switzerland} 
\affiliation[l]{Laboratori Nazionali di Frascati, Via Enrico Fermi 40, 00044 Frascati (Rome), Italy}
\emailAdd{francesco.renga@roma1.infn.it}

\abstract{The reconstruction of the positron trajectory in the MEG-II experiment searching for  the $\mu^+ \to e^+ \gamma$ 
decay uses a cylindrical drift chamber operated with a helium-isobutane gas mixture. 
A stable performance of the detector in terms of its electron drift properties, avalanche multiplication, 
and with a gas mixture of controlled composition and purity has to be provided and  continuously
monitored. In this paper we describe the strategies adopted to meet the requirements imposed by the target 
sensitivity of MEG-II, including the construction and commissioning of a small chamber for an online monitoring of
the gas quality.}


\arxivnumber{...}

\begin{document}
\maketitle
\flushbottom


\section{Introduction}
The MEG-II experiment~\cite{meg2} is under construction at the Paul Scherrer Institut (Villigen, CH) for the search of the 
$\mu^+ \to e^+ \gamma$ decay. MEG-II is an upgrade of the MEG experiment~\cite{meg_det}, 
which collected data from 2009 to 2013 and provided the best upper limit on the branching ratio of this decay, 
$BR(\mu^+ \to e^+ \gamma) < 4.2 \times 10^{-13}$~\cite{meg}. MEG-II is designed to obtain one order of magnitude 
better sensitivity than MEG; the basic idea is to achieve the highest possible sensitivity by making the maximum 
use ($7\times10^{7}$ muons/s) of the available muon intensity at PSI with an improved detector, keeping  the background 
at a manageable level. The detectors used in these experiments should be highly performing, stable and accurately calibrated, 
and several methods have been developed to calibrate and monitor the behavior of all detectors, some of which required 
dedicated beams and/or auxiliary detectors~\cite{CW,TC}. In this paper we consider one of the main ingredients for the 
stability of the drift chamber of the MEG-II positron spectrometer: the delivery of a gas mixture of high and continuously 
monitored quality.

Indeed, the MEG upgrade involves the construction of a cylindrical drift chamber~\cite{cyldch}, 
to replace the 16 drift chamber modules of MEG, which will improve the resolution on the positron 
4-momentum by more than a factor of 2, thanks to a single hit resolution of about 100~$\mu$m in a very light, helium based 
gas mixture. The possibility of reaching such a good resolution has been confirmed by several tests on small prototypes~\cite{proto}. 
On the other hand, it is at the limit of the present technical capabilities, and hence a very stable long-term operation of the detector is 
needed to guarantee that the same performances are obtained in the experiment. In this perspective, the quality of the gas mixture is crucial: 
uncontrolled fluctuations of the gas composition and contaminations by impurities would make the drift velocity unstable, an effect 
hard to be evaluated and corrected by analyzing the muon decay data. 

After a brief introduction to the MEG-II drift chamber (Sec.~\ref{sec:dch}), in Sec.~\ref{sec:gas_system},~\ref{sec:risk} and~\ref{sec:analyzers} 
we will describe the gas distribution system of the MEG-II experiment, including the risk management and the gas analysis strategies which 
will be adopted. In Sec.~\ref{sec:chamber} we will describe the construction and commissioning of a small chamber, which has been
designed to provide a continuous online monitoring of the gas quality during the data taking of the MEG-II experiment. The laboratory tests
performed on this chamber are described
in Sec.~\ref{sec:tests}.

\section{The MEG-II drift chamber}
\label{sec:dch}

The MEG-II drift chamber~\cite{cyldch} is contained in a 2~m-long cylindrical volume filled with square drift cells defined by 8 
silver-plated aluminum cathode wires with a gold-plated tungsten anode wire (20~$\mu$m diameter) at its center. The cells are 
distributed on 10 radial layers of 192 cells each. Cathode wires have a diameter of 40~$\mu$m if they are between two layers
and 50~$\mu$m if they are between two anodes of the same layer. The longitudinal coordinate along the axis of the cylinder ($z$ axis) 
is reconstructed with a stereo geometry arrangement of the wires, slanted alternatively by $\pm8^\circ$ with respect to the $z$ axis. 
Due to these features, the size of the cell changes from about $(6 \times 6)~\mathrm{mm}^2$ in the innermost region of the innermost 
layer to about $(9 \times 9)~\mathrm{mm}^2$ at the end plates of the external layer. Radially, the cells are closely stacked, so that
there are two adjacent layers of cathode wires with opposite stereo angle in between two layers of anode wires. It gives an effective
cathode to anode wire ratio of 5:1. The chamber will be operated with a mixture of helium and isobutane (iC$_{4}$H$_{10}$), in 
a volume proportion of 85\% and 15\% respectively. Tests on prototypes~\cite{proto} show that a spacial resolution of about 
120~$\mu$m can be reached for the coordinate transverse to the $z$ axis.

The chamber is shaped as a tube, with the inner cavity filled with pure helium, so that muons can travel with minimal interactions 
to reach the stopping target placed at the center of the spectrometer. The two volumes are separated by a 20~$\mu$m-thick 
aluminized Mylar \textsuperscript{\textregistered} foil.

\section{Gas Distribution System and Pressure Control}
\label{sec:gas_system}

The purity of the gas mixture is primarily guaranteed by a continuous flow of the gas in the drift chamber volume using 
 helium and isobutane with purity level of 99.9999\% and 99.995\% respectively. All this will prevent the accumulation of 
contaminants originated from material outgassing or formed during the avalanche process. Moreover, this  allows to keep a small 
overpressure (a few Pa) in the chamber with respect to the atmosphere of the experimental hall, even in presence of small leakages, 
in order to avoid contamination by external agents.

\begin{figure}[htbp]
   \centering
   \includegraphics[width=\textwidth]{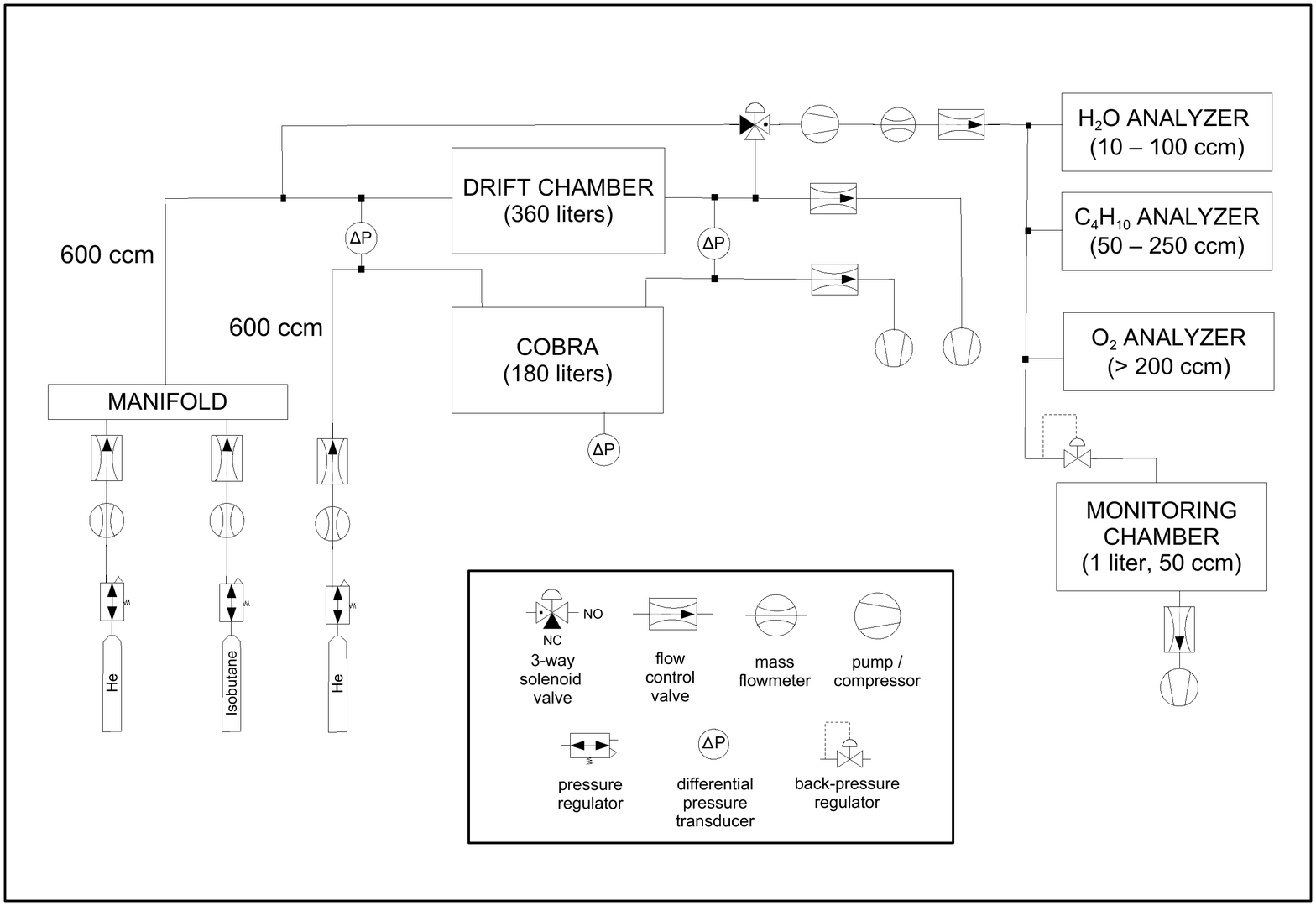}
   \caption{Simplified schematic of the gas system for the spectrometer of the MEG-II experiment. Ancillary lines, valves and monitoring 
   sensors are omitted.}
   \label{fig:gas_scheme}
\end{figure}

A simplified schematic of the system is shown in Figure~\ref{fig:gas_scheme}.
It is primarily used to flow a fresh He:iC$_{4}$H$_{10}$ (85:15) mixture in the drift chamber volume. Moreover, the system also provides a 
continuous flow of pure helium through the inner volume, which is conventionally indicated as the COBRA volume, from the name of the 
magnet surrounding the spectrometer. The gas distribution system is composed of four main sections:

\begin{enumerate}
\item The Gas Supply and Distribution System, which provides a constant and uniformly distributed gas flow to the chamber 
and to the COBRA volume, by means of a set of mass flow controllers (MKS mod. 1179B).\footnote{Information about
MKS products can be found at \url{https://www.mksinst.com}.} In order to have some flexibility in case 
of problems when operating the chamber (e.g. high voltage instabilities or aging), the gas system allows to  add  at least one 
further component (e.g. $\mathrm{CO}_2$) to the chamber admixture and a small fraction of artificial air into COBRA. Hence, the system 
is composed of two pure helium lines (one dedicated to the drift chamber and one to COBRA), one pure isobutane line, one extra line for 
the chamber, one air line for COBRA and one high-rate line which can be alternatively connected to the COBRA inlet or outlet. 
This high rate line is used to compensate for fast pressure changes during the insertion and extraction of an external beam line used
for detector calibration purposes. The distribution system includes a custom-designed 0.7-liter multi-stage manifold for a good mixing 
of the gas components and a system of valves, which are used to release and equalize the pressure in the two volumes in case of 
unsafe conditions. The valves are connected to an alarm system which is implemented, for redundancy, both in software (see below) 
and hardware (double alarm threshold relays with analog inputs from the pressure sensors).
\item The Pressure Control System, which regulates the output flow to maintain constant differential pressures 
between COBRA and the atmosphere, and between the chamber and COBRA. Pressure is measured
by differential capacitance manometers with analog outputs (MKS mod. 226A, $\pm 26.7$~Pa full scale) and regulated by driving 
proportioning valves (MKS mod. 248A) placed at the outlets of the two volumes and connected downstream to two vacuum pumps. 
During the commissioning of the system it was noticed that the absence of a stiff separation between the two volumes makes difficult 
to reach the desired operating conditions if flows are started together in the two volumes. This problem is circumvented by flowing the 
COBRA volume alone until the pressure in the two volumes reaches a stable value.
\item The Gas Monitoring System, which analyzes a sample ($\sim$~10-50\%) of the chamber's output flow (see Sec.~\ref{sec:analyzers}).
\item The Control System, the intelligent part of the apparatus, which drives the hardware components to set the 
flows and regulate the pressures, manages the software-based safety procedures, and is interfaced to the slow control system of the 
experiment. It is based on the Midas Slow Control Bus (MSCB) system and SCS2000 hardware.\footnote{See \url{http://midas.psi.ch/mscb}.}
In particular, a proportional-integral-derivative (PID) control loop is implemented in the SCS2000 for the pressure regulation.
\end{enumerate}


Under normal conditions, the system will be operated with a flow of 600~cm$^3$/min of He:iC$_{4}$H$_{10}$ (85:15) mixture in the
chamber and 600~cm$^{3}$/min of helium in the COBRA volume, equivalent to  a flow of one chamber 
volume and two COBRA volumes in about 10 hours. A maximum flow of about 1500~cm$^3$/min can be delivered by the mass flow
controller, and the possibility of a high flow up to 20 l/min is foreseen with manual flow meters, when a fast cleaning of the volumes
is required. The structural characteristics of the COBRA volume prevent to keep a constant pressure inside the volumes irrespective of the 
atmospheric pressure. Hence, the COBRA volume will be kept at pressure $P_\mathrm{C} = 10$~Pa above the atmospheric pressure. 
The drift chamber volume will be operated at pressure $P_\mathrm{DC} = 5$~Pa below the COBRA pressure,\footnote{The COBRA 
pressure has to be higher that the drift chamber pressure in order to keep smooth the shape of the thin foil separating the two volumes} 
and hence 5~Pa above the external atmosphere. The exhaust gas will not be recycled.

The purity of the gas entering the drift chamber will be also improved by Oxisorb\textsuperscript{\tiny\textregistered} oxygen 
and moisture filters installed on the gas inlets and guaranteeing  contaminations of $< 5$~ppb O$_2$ and $< 30$~ppb H$_2$O.\footnote{See
\url{http://www.spectron.de/spectron_de/en/produkte/spectromol}.}

\section{Risk Management}
\label{sec:risk}

The system has been designed in order to guarantee the safety of both the equipments and the detector in case of anomalous 
conditions or human errors.

The software control is implemented as a state machine. Each machine state is associated in hardware to the state of a set of 
valves and flow controllers, and only a few transitions between states are allowed. In practice, the operator 
cannot maneuver each single valve, but only set of valves into predetermined combinations. It prevents the system from going 
into unexpected configurations due to a human error.

The normal state of the valves is chosen in such a way that, when the system is off or there is a power outage,
the chamber and COBRA volumes are interconnected and open to the atmospheric pressure (so that no stress is applied
to the inner wall), the system is isolated from the gas reservoirs and the pumps are off. The same state is automatically set, 
by alarms implemented for redundancy both in hardware and software, if the differential pressure between the chamber and 
COBRA or COBRA and the atmosphere exceeds $\pm 20$~Pa. It protects the chamber if for any reason the gas system is 
unable to maintain the proper pressures. Similarly, if the gas flow stops or the vacuum pumps cannot guarantee the required 
differential pressure across the pressure control valves, the system goes into a state where the detector volumes are isolated, 
in order to prevent abrupt pressure changes.

\section{Gas Analysis Toolkit}
\label{sec:analyzers}

The most common contaminants which can drastically affect the performances of a drift chamber are oxygen and moisture.  
Electrons easily attach to oxygen, which causes hence a reduction of the charge collected by the anode wires of the chamber.
More critically, variations in the water vapor concentration can modify significantly the drift velocity of electrons,
producing a worsening of the resolution if the moisture level is neither constant nor monitored and corrected for. 

According to our simulations of the performance of the MEG-II spectrometer, a significative deterioration of the detector performances
arises from uncontrolled variations of the drift velocity by more than 5\%. 
Hereinafter we consider variations of 1\% as a benchmark for the evaluation of the performances of the gas analysis and 
monitoring systems.

Detailed studies of the influence of moisture and oxygen in helium-isobutane mixtures have been performed in the past~
\cite{grancagnolo}. According to these results, 1\% effects on the drift velocity can be produced by
moisture variations as low as 1000 ppm. It is also known that water enhances the electron attachment to oxygen~\cite{huk}.
Oxygen contaminations are less critical, because in MEG-II we do not expect to extract any information from the measured 
collected charge. Nonetheless, it is important to monitor the oxygen content of the mixture, in order to avoid any misinterpretation 
of sudden gain losses and to detect possible leakages that cause air entering the system.

For these reasons, a trace oxygen analyzer and a moisture analyzer will be installed in the gas system of the MEG-II experiment.
The oxygen analyzer is a Teledyne 3000TA,\footnote{Information about products by Teledyne Analytical Instruments can be 
found at \url{http://www.teledyne-ai.com}.} with a programmable range from 0-10~ppm to 0-250000~ppm and an accuracy
of $\pm 2\%$ of the full scale. The moisture analyzer is a MEECO Accupoint LP2\texttrademark,\footnote{See \url{http://meeco.com}.} 
with a 0-5000~ppm range and an accuracy of $\pm 5\%$ of the reading. According to the discussion above, these devices are fully 
compliant with our requirements.

Another critical feature is the stability of the gas mixture composition. According to simulations performed with the GARFIELD
software,\footnote{See \url{http://garfield.web.cern.ch/garfield}.} in order to keep the drift velocity stable within 1\% the isobutane 
fraction has to be stable at $(15.0 \pm 0.5)\%$, which means a precision of about 3\%. The flow controllers installed in the gas 
distribution system are expected to guarantee a precision of 0.1\%. On the other hand, pure helium can permeate through the thin 
foil separating the chamber volume from the COBRA volume, and the composition of the mixture in the chamber can be different with 
respect to the one delivered by the gas supply system. Hence, an infrared gas analyzer (Teledyne 7300A) will be installed, with 
the capability of measuring the isobutane content of the mixture in the range 0-30\%, with a repeatability of $\pm 1\%$ of the full 
scale. It will allow to tune the composition of the fed-in mixture in order to reach an equilibrium where the composition inside the 
chamber is the required one.

\section{Gas Monitoring Chamber}
\label{sec:chamber}

Gas analyzers can detect only a few kinds of contaminations with the required precision. Moreover, it can be difficult
to determine the real impact of a generic contamination on the performances of the detector. Hence, in order to have a more
general and continuous monitoring of the quality of the gas used in the experiment, we plan to install a small drift chamber 
in the gas distribution system. The basic idea is that a small chamber, with a very simple geometry,
allows to detect instabilities of the relevant gas properties in a  quicker and  more reliable way with respect to the 
large drift chamber installed in the experiment. Moreover, if the monitoring chamber can be supplied with gas coming
alternatively from the inlet or outlet of the detector, it is possible to determine if the contamination originates inside the chamber 
or in the gas supply system (including a possible bad quality of the supplied gas).

The monitoring chamber has been primarily designed to perform a quick and frequent measurement of the
multiplication factor (\emph{gain}) $G$ in electron avalanches produced by X-rays from a radioactive source.
We also investigated the possibility of monitoring the drift velocity with the same detector using cosmic rays.

The contamination by electronegative molecules, which increase the electron attachment probability, can be detected with the gain
measurement. Although this parameter is not of great relevance for the MEG-II drift chamber, this technique can provide a fast alert in case 
of a sudden contamination which can potentially impact also the drift properties, and hence the performances of the detector. 
Conversely, monitoring the drift velocity allows to detect contaminations which have a direct impact on the performances of the MEG-II drift
chamber, and possibly to correct for their effects. 

The gain will be monitored by measuring the charge collected by a single wire when a 5.9~keV X-ray from a $^{55}$Fe 
source ionizes  one  gas molecule, producing a delta ray with approximately the same energy of the incoming X-ray.
The delta ray looses all its energy within a few hundred microns, by further ionizing the gas. According to simulations performed
with the DEGRAD software,\footnote{See \url{http://consult.cern.ch/writeup/magboltz}.} about 250 electrons are expected to be produced by
the interaction of a single X-ray. In a drift cell, the electrons drift toward the anode wire and produce an avalanche. The total charge can 
be measured by integrating the electronic signal induced on the wire or measuring its  amplitude at the maximum.

The drift velocity measurement will be performed by reconstructing tracks produced by charged particles going through many
cells in the detector. Signals on all cells but one can be used to reconstruct tracks and their distance from the anode wire of the  excluded cell,
where the signal time is measured. The drift time-to-distance relation is derived, and hence the drift velocity.

\begin{figure}[htbp]
   \centering
   \includegraphics[angle=-90,width=0.45\textwidth]{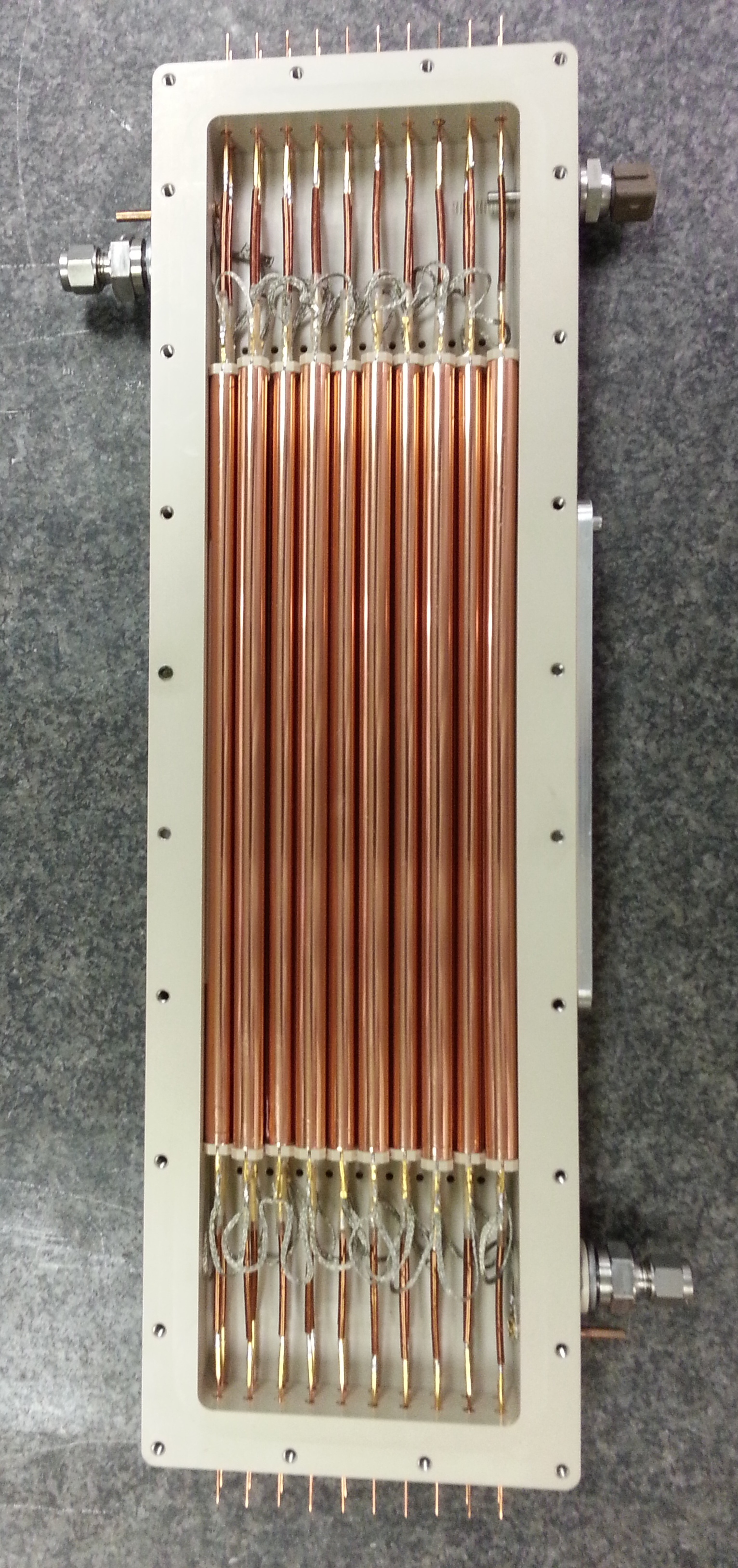} 
   \hspace{0.5cm}
   \includegraphics[angle=-90,width=0.45\textwidth]{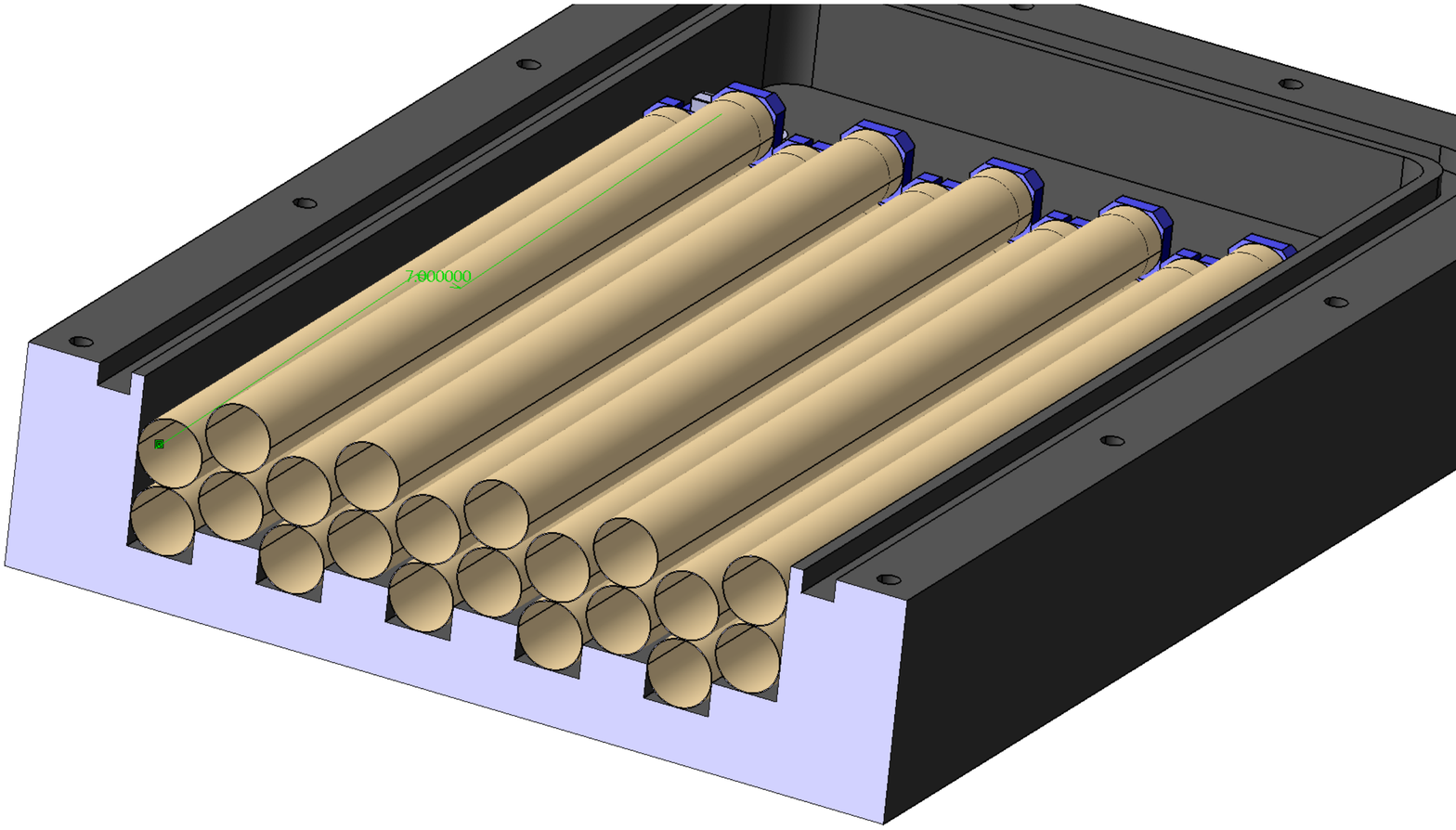} 
   \caption{A picture of the assembled monitoring chamber before closing the gas box (left) and a drawing of the transverse
   section of the detector (right).}
   \label{fig:chamber}
\end{figure}

The design of the monitoring chamber had to satisfy three requirements: a low material budget,  to allow X-rays to penetrate
inside the active volume and to minimize the multiple Coulomb scattering (MS) experienced by charged particles; a very simple drift cell
geometry for  a precise determination of the time-to-distance relation; a sufficiently large number of cells 
to have a good track reconstruction. Hence, we decided to build a chamber made of 20 thin-wall drift tubes. The tubes are disposed in 10 
columns, with one pair of tubes per column (see Figure~\ref{fig:chamber}). The very simple and very symmetric geometry of a drift tube is 
the main advantage of this design, because the measurement of the drift velocity is not affected by the impinging particle angle 
in the chamber.

Each tube is made of a 36~$\mu$m-thick  copper-coated Mylar wall kept at ground voltage, with a diameter of 9.75~mm and a 
length of 25~cm. On the axis of the tube a 20~$\mu$m diameter gold-plated tungsten wire is placed. The wire 
is kept in position by two caps at the end of the tube. The cap is made of PEEK, with a copper pin at the center where the wire 
is soldered and the front-end electronics and high voltage connections are made (only 16 tubes are instrumented).
A copper ring at the interface with the tube has been inserted for grounding connections. Each pair of tubes is staggered with 
respect to the nearby one by half a tube diameter. 

Instead of having single gas-tight tubes, all tubes are placed within a single gas-tight box, also made of PEEK, and the caps have
apertures to let the gas flow through them. High voltage is provided through two printed
circuit boards, placed out of the box, with protection resistors and an RC low-pass filter. Front-end preamplifiers, having the
same design of the ones used in the MEG-II drift chamber~\cite{meg2}, are also placed out of the box. High voltage 
connections inside the box are made with kapton-isolated copper wires, in order to guarantee good dielectric strength and
low gas contamination by outgassing. Finally, the box has a thin Kapton window to allow  X-rays from an external source to enter
the box. A PT100 sensor is also installed in the chamber and it is used to monitor the temperature of the gas inside the
detector. Figure~\ref{fig:chamber} shows a transverse section of the assembly. The chamber, including
the front-end electronics and the high voltage distribution boards, is placed inside an aluminum box for electromagnetic shielding.

The design of the chamber has been validated with calculations and simulations performed with 
GEANT4~\cite{geant4}, GARFIELD and DEGRAD. The results show that:
\begin{enumerate}

\item An X-ray source placed at a distance of 2~cm from the Kapton window should have an activity of about 20~MBq in order
to produce an event rate of about 1~kHz in the first tubes crossed by the X-rays (such a high event rate is required in order 
not to be overwhelmed by electronic noise events observed in previous prototypes with a rate up to several 10~Hz).
\item MS by cosmic muons would contribute negligibly to the tracking performances. 
\item The tracking resolution contributes negligibly to the measurement of the time-to-distance relation if the
number of tubes in a track is larger than 6.
\item The acceptance of the chamber for cosmic rays allows to collect about 20000 good tracks in one week of continuous data
taking.
\end{enumerate}

\section{Tests of the monitoring chamber}
\label{sec:tests}

In this section we will describe the tests we performed in order to assess the performances of the monitoring chamber.

\subsection{Tests of the gain monitoring capabilities}

The monitoring of the electron amplification factor $G$  is affected by the atmospheric conditions (pressure and temperature) according to
the Diethorn formula~\cite{diethorn,blum_rolandi}, which predicts the relative variations of the gain to be proportional to the
relative variations of the gas density,  $dG/G \propto -d\rho/\rho$. Hence, we expect $G$  to increase with decreasing
pressure and increasing temperature. If the chamber is not operated at fixed temperature and pressure, as it could be in
our application, changes in atmospheric conditions could fake or mask a gain variation due to gas contaminations. Hence,
we performed a set of measurements at fixed pressure and temperature in order to determine how to correct for these
effects. In this test, the pressure was controlled and monitored by a dedicated gas system, while the temperature was
measured by the PT100 sensor installed in the chamber. We were able to change the temperature of the gas by regulating the
room temperature and the ventilation of the electronics in the gas system, although it only allowed for small, slow and uneven 
variations.

\begin{figure}[htbp]
   \centering
   \includegraphics[width=0.45\textwidth]{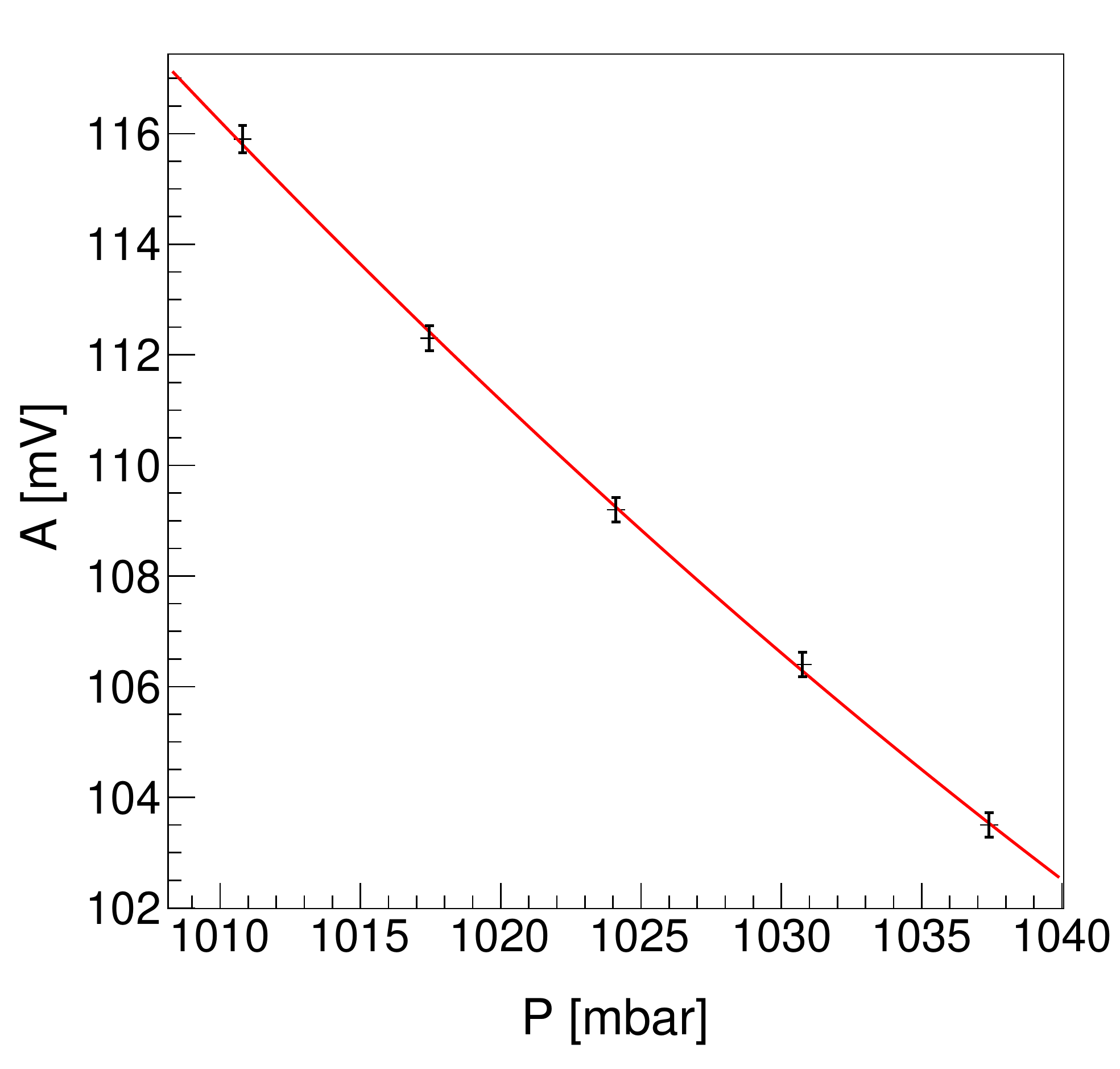} 
   \includegraphics[width=0.45\textwidth]{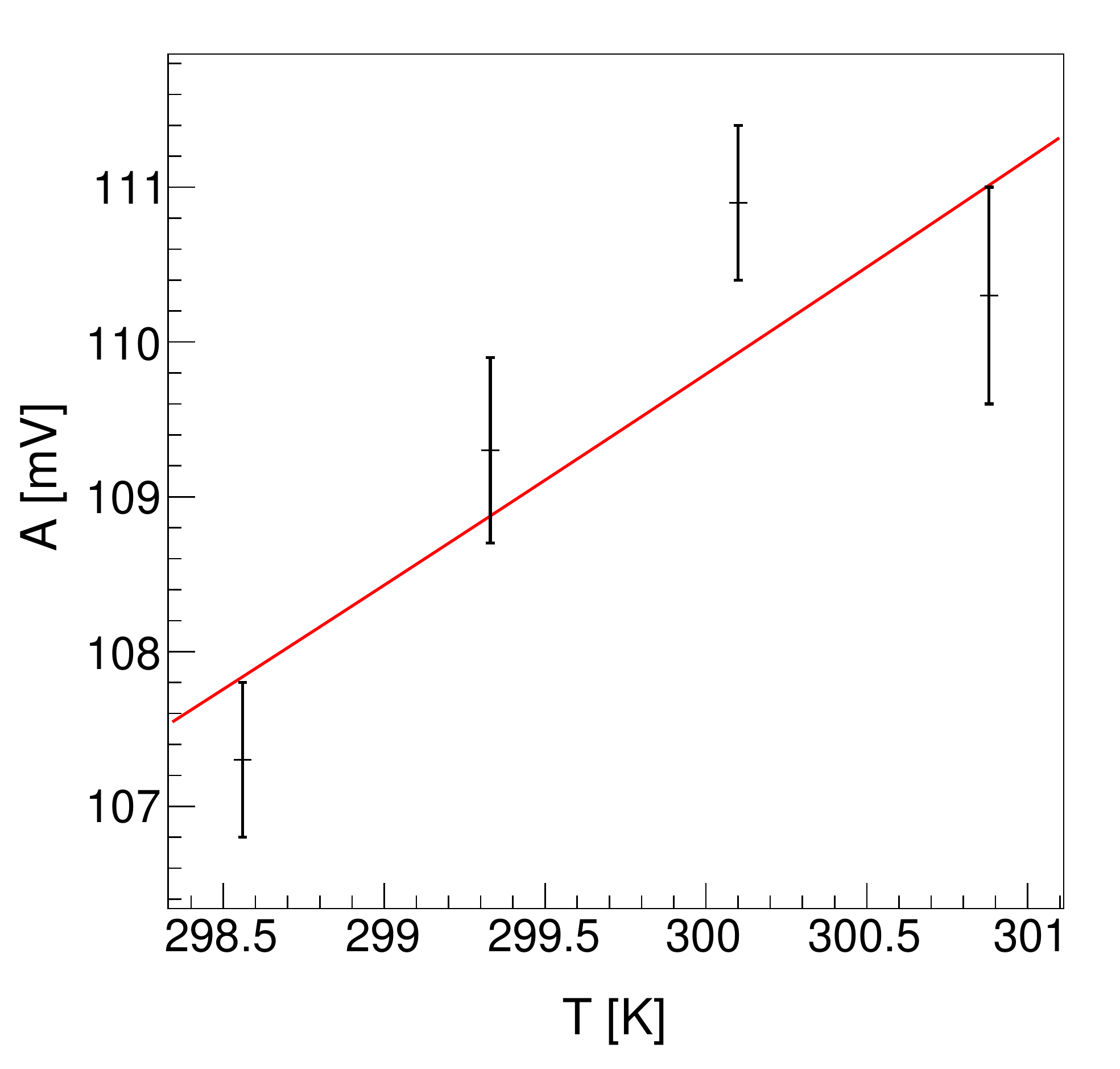} 
   \caption{Average amplitude $A$ of signals produced by 5.9~keV X-rays from a $^{55}$Fe radioactive source, as a function
   of the pressure at a fixed temperature of $299.3$~K (left) and as a function of the temperature at a fixed pressure of 
   1027~mbar (right). The measurements are fitted by a second-order Taylor expansion of the Diethorn formula 
   $\frac{dA}{A} = -\frac{V \log2}{\Delta V\log(b/a)}\frac{d\rho}{\rho}$, where $a$ and $b$ are the wire and tube radii, V is the anode
   potential and $\Delta V$ is an empiric parameter which characterizes the gas mixture.}
   \label{fig:dG}
\end{figure}

Figure~\ref{fig:dG} shows the average amplitude of signals produced by X-rays as a function of pressure (at a fixed temperature
of $299.3$~K) and temperature (at a fixed pressure of $1027$~mbar). The chamber was operated with a voltage of
1600~V on the anode wires, where a gain of a few $10^5$ is expected from GARFIELD simulations, similar to what is expected
in the MEG-II chamber. The measurements at fixed temperature have been repeated four times, in order to verify 
their stability. The results are consistent over the four sets, and the average of the four measurements is shown here.
The expected trends are observed and  the parameter $\Delta V$ in the Diethorn formula, 
$\Delta V = (41.5 \pm 1.0)$~V, is derived. 

\begin{figure}[htbp]
   \centering
   \includegraphics[width=0.45\textwidth]{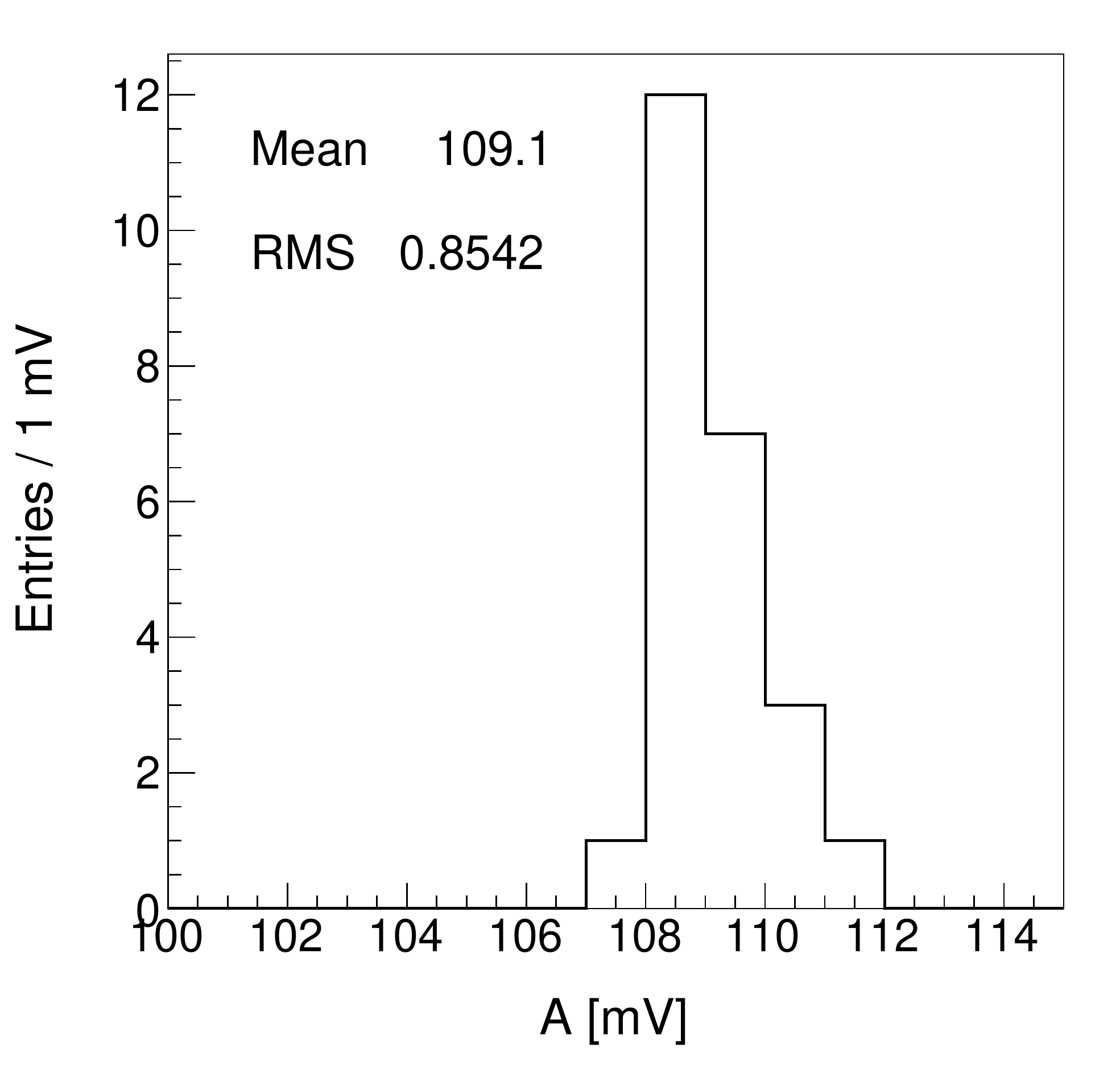} 
   \caption{Distribution of the average amplitude of signals produced by 5.9~keV X-rays from a $^{55}$Fe radioactive source,
   on data sets collected under different atmospheric conditions and corrected to a reference working point, 
    $P = 1027$~mbar and $T = 299.3$~K.}
   \label{fig:gain_corr}
\end{figure}

These trends allow to correct for variations of pressure and temperature. In Figure~\ref{fig:gain_corr} all measured values
are corrected to a reference working point, $P = 1027$~mbar and $T = 299.3$~K. 
The width of the distribution gives the precision of the gain measurement, including our ability of correcting for variations of pressure 
and temperature. A precision of $\sim 1\%$ is reached, which corresponds to 0.6 K temperature variations (at constant pressure) and
2 mbar pressure variations (at constant temperature). As a measurement of the goodness of this performance for the MEG-II goals, 
one can consider that, according to our measurements and simulations, a 3\% change in the Isobutane concentration (0.45\% over 15\%)
gives 1\% effects on the drift velocity and 5\% effects on the gain.

A measurement of the gain drop induced by an increase of the isobutane fraction has been also performed. After having measured
the average amplitude for a given pressure and temperature working point, the inlet gas flows were changed to increase the Isobutane
fraction from 15 to 20\%, with a total flow of 50~cm$^3$/min. The signal amplitude was monitored at regular time intervals and, 
after about 1 hour, it reached the new stable value, $(50.4 \pm 0.3)\%$ of the original one.


\subsection{Tests of the drift velocity monitoring}

We collected data at the Beam Test Facility (BTF~\footnote{See \url{http://www.lnf.infn.it/acceleratori/btf}.}) of the INFN Laboratori 
Nazionali di Frascati to test the tracking performances of the monitoring chamber under different conditions. 

\begin{figure}[htbp]
   \centering
   \includegraphics[width=0.7\textwidth]{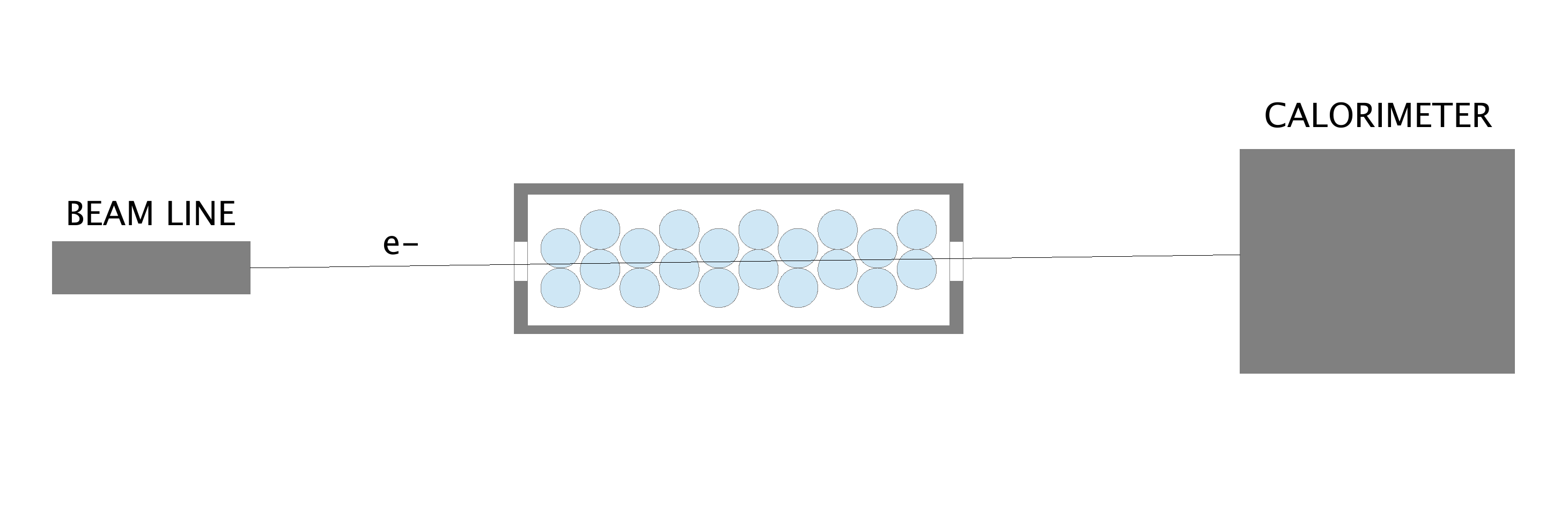} 
   \caption{Schematic view of the BTF setup (not in scale).}
   \label{fig:BTF}
\end{figure}

The BTF provides an electron beam, with an energy of about 450~MeV. A timing signal of the LINAC in coincidence 
with the delivery of a bunch at the BTF determines the arrival time of the electrons at the experimental area up to a constant
offset. The beam was tuned to get about one electron per spill on average. Electrons cross the chamber, making tracks in a row of tubes, 
and then impinge on a calorimeter that allows to count the number of electrons in a single spill. Events with only one electron are selected 
offline. A schematic view of the setup in shown in Figure~\ref{fig:BTF}.

The temperature was monitored and found to be stable during the data taking, while the pressure was controlled by a dedicated gas 
system. The data collected at 1000 mbar and 1625 V provided about 10000 tracks which could be successfully reconstructed with a 
signal in at least 4 tubes and $\chi^2 < 30$. The drift time is measured as the difference between the leading edge of the signal in a drift 
tube and the time of the track given by the LINAC signal, after subtracting the constant offset, which can be measured by requiring 
that the distribution of the drift times start from zero (corresponding to particles passing very near to the wire). 

A first tracking iteration is performed assuming a constant drift velocity. It allows to convert the drift time into a distance $d$ 
from the wire, and to fit the track whose distances from the wires give the better agreement with the measurements. Tracks
reconstructed in this way are then used to build a time-to-distance relation, by looking at the measured drift times in a wire as a
function of the distances of the fitted tracks from that wire. This relation can be used to iterate the procedure and get a better
estimate of the relation itself.

\begin{figure}[htbp]
   \centering
   \includegraphics[width=0.45\textwidth]{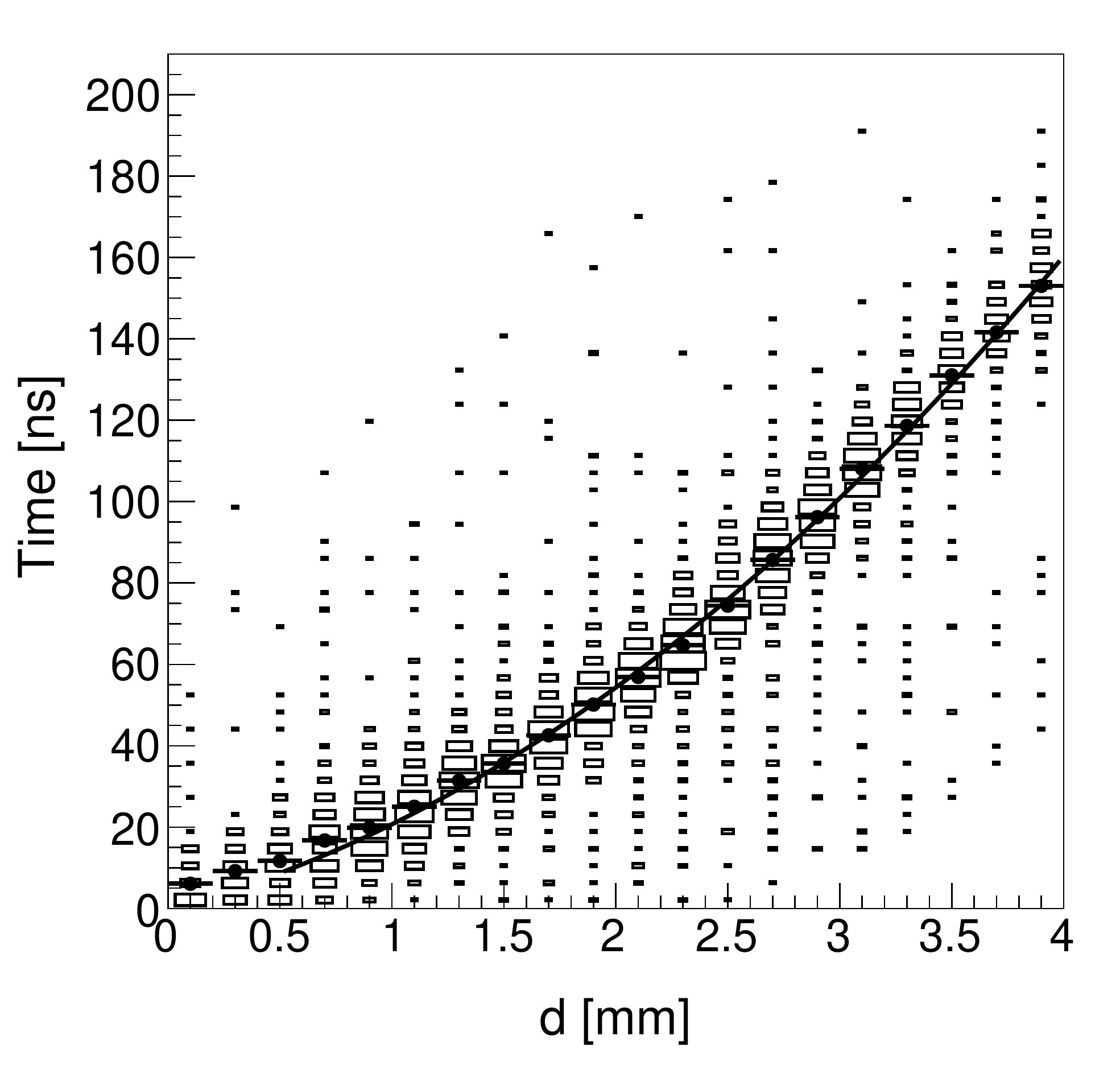} 
   \caption{Time-to-distance relation measured at the BTF with the monitoring chamber operated at 1625~V and 1000~mbar.}
   \label{fig:TtoD_ex}
\end{figure}

Figure~\ref{fig:TtoD_ex} shows an example of a time-to-distance relation, fitted with an order 2 polynomial.
An analysis of the uncertainties shows that the benchmark sensitivity of 1\% would require, with cosmic rays, 
an unsatisfactorily long time of about two weeks. We are investigating possible alternatives to have a much 
faster monitoring of the drift velocity using radioactive sources, with this or an ad hoc device.

\section{Conclusions}

A good and continuously monitored gas quality is necessary in order to reach the design performances of the drift
chamber of the MEG-II experiment. It will be guaranteed by a gas distribution system that includes a set of gas analyzers, 
to monitor the most important contaminants and the composition of the mixture, and a monitoring drift chamber which performs
measurements of the gas gain. 

We presented in this paper the requirements and the design of the system, along with a discussion of the performances of the 
monitoring chamber. Tests performed with this detector show that the gain variations can be monitored with percent precision 
with a $^{55}$Fe radioactive source. Although the gas gain is not a critical variable in the MEG-II spectrometer, this technique 
could allow to detect abrupt variations of the gas composition due to contaminants. We also investigated the possibility of
monitoring the drift velocity with the same device and using cosmic rays. A 2\% precision can be reached in a time 
scale of about three days, which is not completely satisfactory. Alternative solutions for a faster monitoring of the drift
velocity will be considered in the future.

\section*{Acknowledgements}
We are grateful for the support and co-operation provided 
by Riccardo Lunadei, Daniele Ruggieri and Antonino Zullo from INFN Roma
technical services.

\bibliography{mybibfile_jinst}

\end{document}